# Au and Ag nanoparticles produced by ion implantation in single-crystalline $\beta$-Ga$_2$O$_3$

D. M. Esteves[1,2,*,#], A. S. Sousa[1,2,*,#], I. Freitas[1], Â. R. G. Costa[3], J. Madureira[3,4], S. Cabo Verde[3,4], K. Lorenz[1,2,4], M. Peres[1,2,4]

[1]  INESC Microsystems and Nanotechnology, Rua Alves Redol 9, Lisbon 1000-029, Portugal

[2]  Institute for Plasmas and Nuclear Fusion, Instituto Superior Técnico, University of Lisbon, Av. Rovisco Pais 1, Lisbon 1049-001, Portugal

[3]  Centre for Nuclear Sciences and Technologies, Instituto Superior Técnico, University of Lisbon, Estrada Nacional 10, km 139.7, Bobadela 2695-066, Portugal

[4]  Department of Nuclear Science and Engineering, Instituto Superior Técnico, University of Lisbon, Estrada Nacional 10, km 139.7, Bobadela 2695-066, Portugal

\*  Corresponding authors: duarte.esteves@tecnico.ulisboa.pt; ana.sofia.sousa@tecnico.ulisboa.pt

\#  These authors contributed equally to this work.

**Abstract:** This work reports the successful formation of Ag and Au nanoparticles in $\beta$-Ga$_2$O$_3$ single-crystals by ion implantation and annealing at 550 °C. X-ray diffraction measurements revealed that nanoparticles were formed after the annealing step, presenting a highly-ordered crystalline structure and conforming to a crystallographic relation with respect to the matrix: $(0\bar{1}0)_\beta \parallel (110)_{\text{Ag/Au}}$ and $[102]_\beta \parallel [1\bar{1}2]_{\text{Ag/Au}}$. The presence of these nanoparticles was also confirmed via absorbance measurements revealing the localised surface plasmon resonance peaks associated with these particles. Considering the multiple advantages and the versatility of metallic nanoparticles, their combination with the exceptional properties of $\beta$-Ga$_2$O$_3$ paves the way for a wide range of applications.

1. Introduction

$\beta$-Ga$_2$O$_3$ is an emerging ultra-wide bandgap semiconductor material with promising future applications due to its interesting properties. Its Baliga figure of merit is larger than that of other wide bandgap semiconductors such as GaN or SiC [1], profiting from its wide bandgap of ~4.9 eV at room temperature and large breakdown electric field of ~8 MV/cm. Concerning its optical properties, although studies addressing its nonlinear response remain limited, recent reports highlight its strong potential for integrated photonic applications, particularly in the ultraviolet (UV) and visible spectral regions [2]. This potential arises from its wide transparency window, compatibility with the III-nitride material platforms [3], and comparatively reduced nonlinear coefficients, including a two-photon absorption coefficient approximately 20 times lower and a Kerr nonlinear refractive index 4–5 times smaller than those of GaN [2]. These properties drive applications in the realms of high-power electronics [4] and optoelectronic devices [5], including solar-blind photodetectors [6–8], ultra-low loss waveguides and resonators [2], among others.

These outstanding optoelectronic features make $\beta$-Ga$_2$O$_3$ an excellent candidate for the integration of plasmonic nanostructures, such as metallic nanoparticles (NPs), that could provide a versatile platform to explore both plasmon-induced hot-carrier generation and enhanced nonlinear optical properties. Owing to its wide bandgap and broad transparency window in the visible and near-infrared regions, $\beta$-Ga$_2$O$_3$ minimizes competing linear absorption while enabling efficient excitation of localized surface plasmon resonances (LSPRs) in embedded NPs. The non-radiative decay of these plasmons can generate energetic hot electrons capable of being injected across metal/semiconductor interfaces, enabling sub-bandgap multispectral [9–11] and polarisation-sensitive photodetection [9,12–14]. Moreover, the strong local electromagnetic field enhancement associated with the LSPR can significantly amplify third-order nonlinear effects in the host matrix, which are very interesting for pulsed laser generation [15]. The combination of these mechanisms positions plasmonic nanoparticle $\beta$-Ga$_2$O$_3$ composites as promising candidates for multifunctional photonic and optoelectronic devices operating in the UV–visible spectral range.

Ion implantation presents itself as an elegant solution for the creation of these NPs, as previously demonstrated in materials such as TiO$_2$, SiO$_2$, ZnO or MgO [16–20]. This is a powerful and versatile method that allows for a precise spatial control in the placement of the ions [21], overcoming solubility limits, and in a way that is compatible with a myriad of materials, including both thin films and single-crystals [22]. A key factor to tune the plasmonic properties of the produced nanoparticles is the ability to precisely control their size, morphology, shape and interparticle distance [23–25]. Additionally, the monoclinic lattice of this material presents intriguing behaviours upon ion irradiation and implantation, such a disorder-induced phase transformation to its defective spinel (cubic) $\gamma$-phase [26,27] and anisotropic elastic response [28].

In this work, we demonstrate the synthesis of plasmonic Ag and Au nanoparticles in single-crystalline $\beta$-Ga$_2$O$_3$ via ion implantation and investigate their crystallographic relationship with the host lattice, taking the expected $\beta$-to-$\gamma$ phase transformation of the matrix into account.

2. Results and discussion

Two samples with a (010) surface orientation were implanted with fluences of $5.0 \times 10^{16}$ cm$^{-2}$ of either Ag$^+$ or Au$^+$ ions with energies of 110 and 160 keV, respectively. The energies were chosen in order to have similar projected ranges inside the material — about 30 nm, according to Monte Carlo simulations performed using the Stopping and Ranges of Ions in Matter (SRIM) code [29]. The beam was tilted by 10° with respect to the surface normal in order to avoid the channelling effect. Subsequently, these samples were annealed at 550 °C in air during 30 min in a tube furnace.

X-ray diffraction (XRD) measurements were performed in order to assess the structural properties of the sample after the implantation and after annealing. Figs. 1a and 1b show the symmetric $2\vartheta$-$\omega$ scans obtained about the 020 reflection of $\beta$-Ga$_2$O$_3$ before and after annealing, for Ag- and Au-implanted samples respectively.

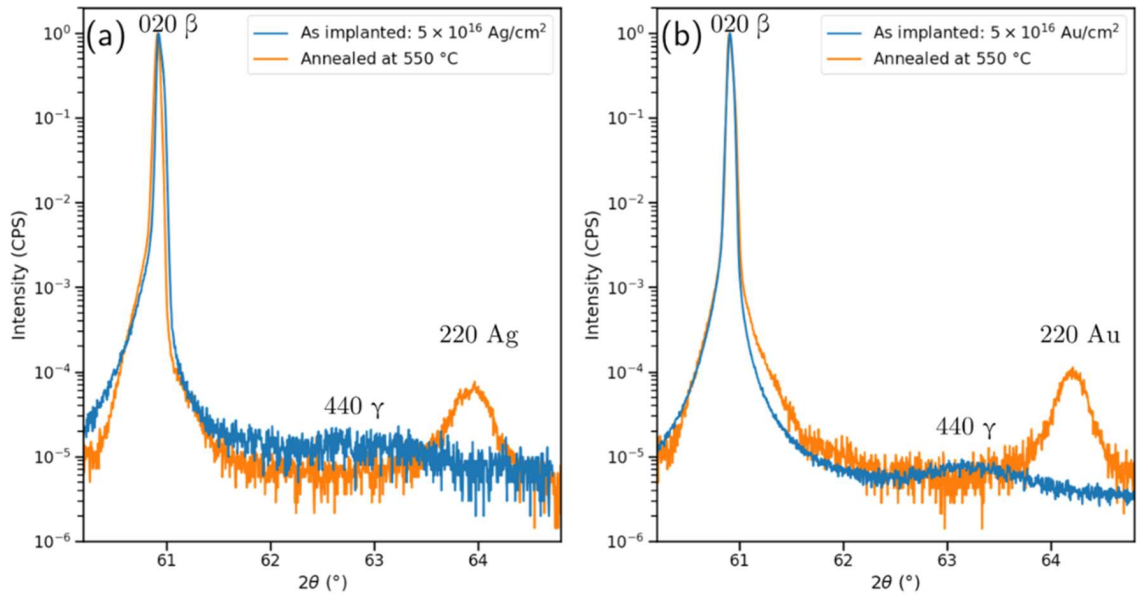

Fig. 1 | Symmetric $2\vartheta$-$\omega$ scans measured about the 020 reflection of $\beta$-Ga$_2$O$_3$ for samples implanted with either (a) Ag or (b) Au, before and after annealing.

In both samples, in addition to the intense sharp peak associated with the expected 020 reflection of $\beta$-Ga$_2$O$_3$, a broad peak appears at about 63° after implantation, and can be assigned to the 440 reflection of $\gamma$-Ga$_2$O$_3$. This observation agrees with previous works showing that this phase can be induced by the disorder introduced by ion implantation [26]. The simultaneous observation of these two reflections in the symmetric scan implies that the {110} family of planes of the induced $\gamma$-Ga$_2$O$_3$ is parallel to the (010) plane of the $\beta$-Ga$_2$O$_3$ matrix. Remarkably, after annealing, the peaks associated with the $\gamma$-phase vanish, while pronounced peaks appear at about 64°, which is close to the 220 reflection peak for single-crystalline Ag and Au [30]. It is also interesting to note that the $2\vartheta$ value for Au is greater than that of Ag, in agreement with the fact that the lattice constant of the former is smaller than that of the latter ($a_{Ag} = 4.10$ Å [31] and $a_{Au} = 4.07$ Å [32]). In the current work, the position of the peaks for Ag and Au is, respectively, $2\theta_{Ag} = 63.93°$ and $2\theta_{Au} = 64.20°$ (obtained via fitting the peaks to a Voigt function), corresponding to lattice constants of $a_{Ag} = 4.12$ Å and $a_{Au} = 4.10$ Å, which suggests the presence of a tensile strain. Since these peaks are quite well-defined, this result suggests the precipitation and subsequent nucleation of the implanted ions, resulting in the formation of crystalline nanoparticles. This process has been observed before in other amorphous and crystalline materials when the concentration of implanted ions surpasses the solubility limit [16–20].

In this context, Fig. 2 shows the reciprocal space maps (RSM) obtained about the 020 reflection in the region containing both the peaks assigned to the matrix and to the Ag and Au nanoparticles, as measured for the annealed samples. Note that, in this representation, the $Q_\parallel = 0$ nm$^{-1}$ line corresponds to a symmetric $2\vartheta$-$\omega$ scan, as the one shown in Fig. 1. The presence of peaks with well-defined shapes corroborates that the Au and Ag nanoparticles must be formed in a highly-oriented manner, suggesting the possibility of a strong influence of the crystallographic orientation of the $\beta$-Ga$_2$O$_3$ matrix.

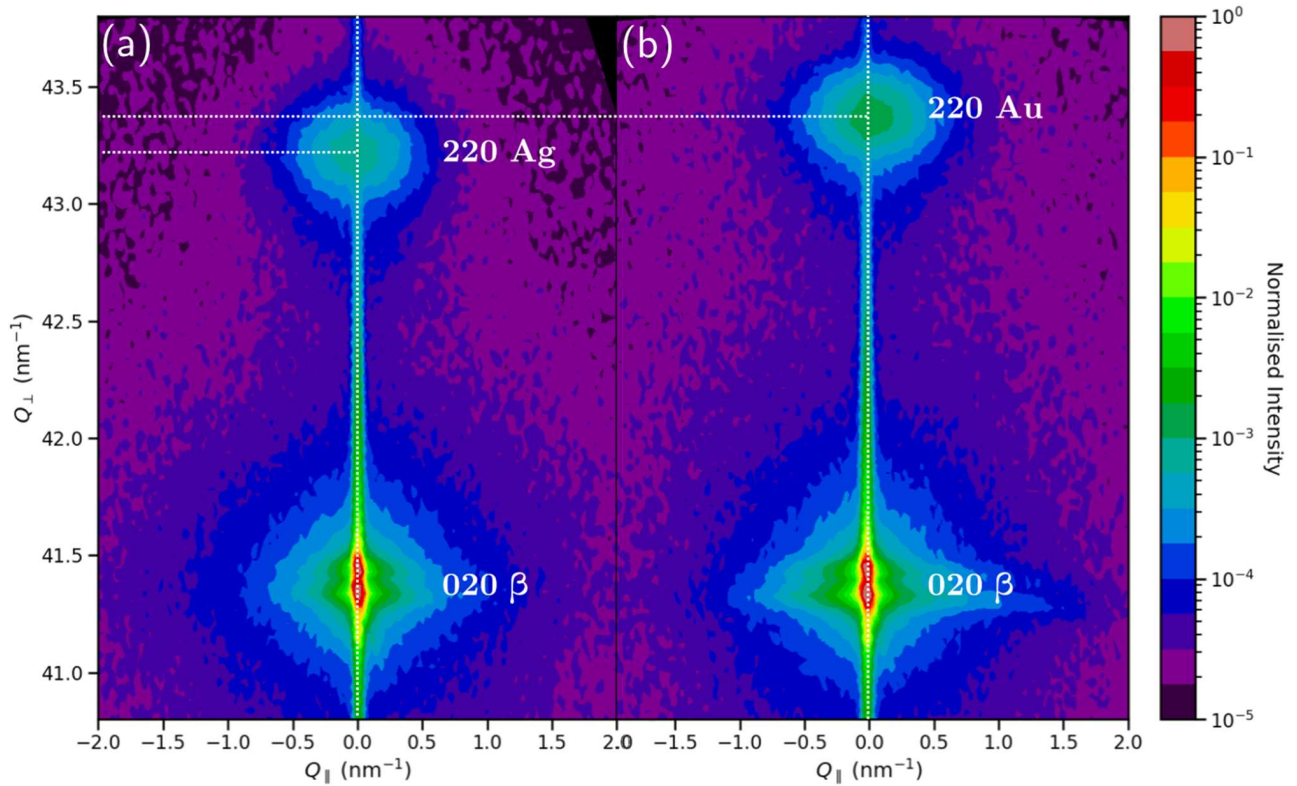

Fig. 2 | Reciprocal space maps obtained about the 020 reflection of $\beta$-Ga$_2$O$_3$ for samples implanted with either (a) Ag or (b) Au. The marked $Q_\parallel = 0$ nm$^{-1}$ line corresponds to a symmetric $2\vartheta$-$\omega$ scan, while $Q_\perp$ is inversely proportional to the lattice constant $a$ for the Ag/Au nanoparticles and $b$ for the matrix.

In order to assess the relative orientation of the nanoparticles with respect to the matrix, pole figures were measured at a detection angle of $2\vartheta = 64°$, thus probing the family of reflections which are crystallographically equivalent to 220 in both the Au or Ag nanoparticles. Figs. 3a and 3b show the pole figures for the samples implanted with Ag and Au, respectively. These figures clearly show the presence of well-defined, sharp peaks pertaining to the pristine matrix, which are superimposed on broader and less-intense peaks attributed to the nanoparticles. This is made evident in Fig. 3c, where the signal was averaged over the tilt angle $\chi$ and shown as a function of the azimuthal angle $\varphi$. This figure thus reveals that there is a very well-defined crystallographic relation between the host (sharper peaks) and the nanoparticles (broader peaks), which can be written as $(0\bar{1}0)_\beta \parallel (110)_{Ag/Au}$ and $[102]_\beta \parallel [1\bar{1}2]_{Ag/Au}$. Notably, this is exactly the same epitaxial relation found between the $\beta$- and $\gamma$-phases when the latter is induced in the former by ion implantation [27,28]. The crystalline structure of these two polymorphs are intimately related, as they share their O sublattices. Remarkably, the lattice parameter of the cubic $\gamma$-phase ($a_\gamma = 8.22$ Å [33]) is almost exactly twice that of Ag and Au single-crystals ($2a_{Ag} = 8.20$ Å [31] and $2a_{Au} = 8.14$ Å [32]), and is related with the $b$ lattice parameter of the $\beta$-phase by the approximate relation $a_\gamma \sim 2\sqrt{2}b$. All these similarities are evidenced by the fact that the 020 $\beta$ peak, the 440 $\gamma$ peak and the 220 Au/Ag peaks lie in the vicinity of one another, as shown in Fig. 1. This observation suggests that the crystalline structure of $\beta$-Ga$_2$O$_3$ may be particularly interesting to accommodate the formation of these nanoparticles in a highly-organised manner, in a situation akin to domain-matching epitaxy [34].

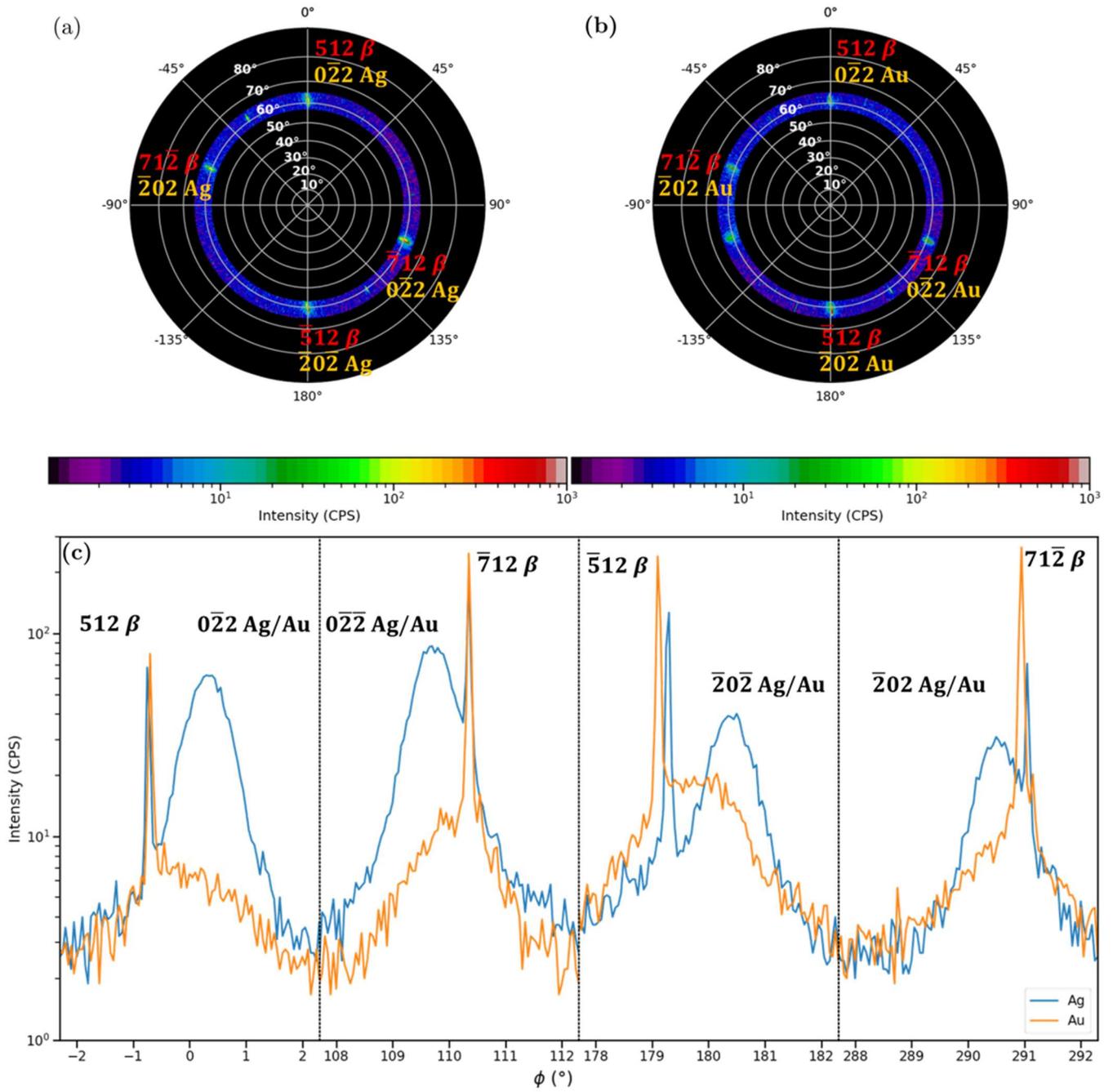

Fig. 3 | Pole figures obtained for the {110} family of planes of (a) Ag and (b) Au. Panel (c) shows the signal as a function of the azimuthal angle $\varphi$ upon averaging over the tilt angle $\chi$.

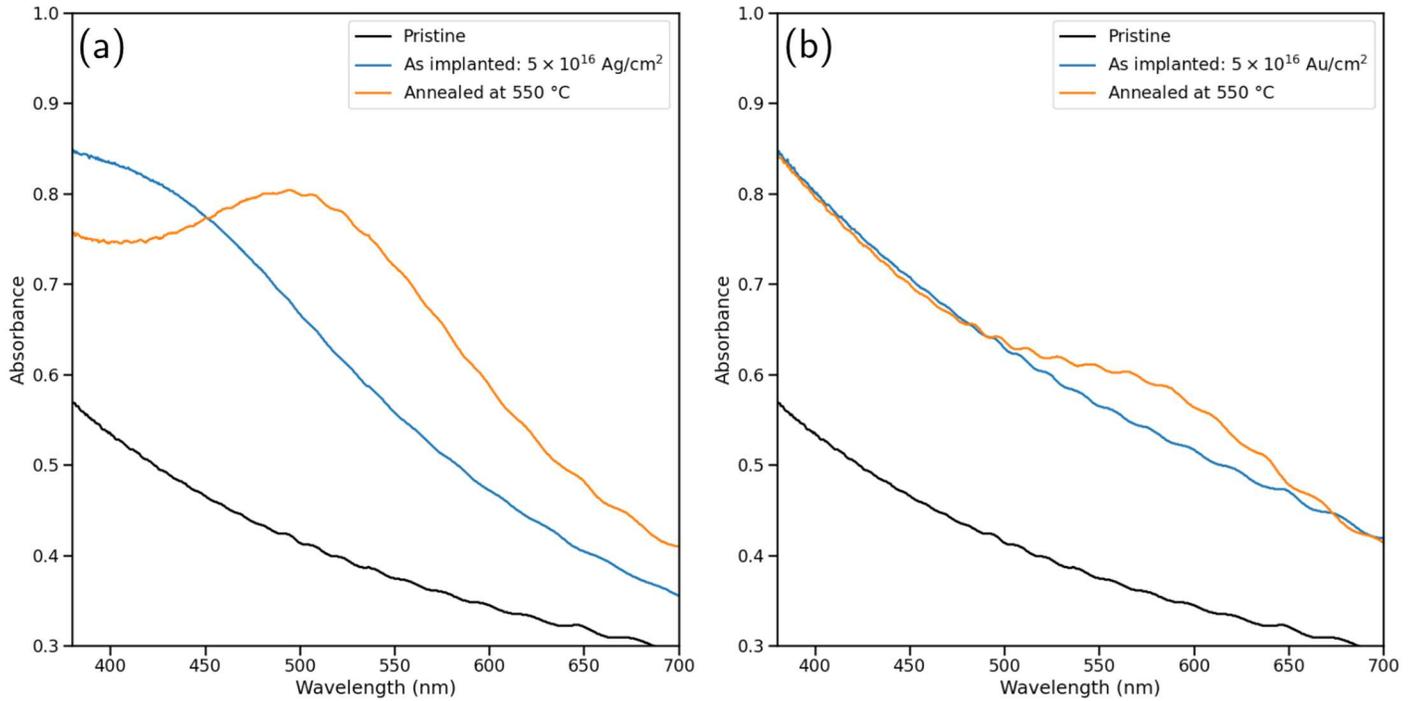

Fig. 4 | Absorbance spectra in the visible region for the sample implanted with either (a) Ag or (b) Au, after implantation and after annealing. The spectrum for a pristine sample is also shown as a reference.

Fig. 4 shows the visible absorbance spectra of each sample after implantation and after annealing, as well as a spectrum of a pristine sample. In the case of the pristine samples, the low absorption in the visible region confirms the good transparency of the material. The absorption spectra of Ag- and Au-implanted samples is enhanced in the visible region, which may be related with the defects introduced during ion implantation. In the case of the Ag-implanted sample, it is possible to notice a weak and broad absorption band around 400 nm already in the as implanted spectrum, while no well-structured features are observed for the Au-implanted sample. However, after the annealing, we observe absorption bands centred at approximately 500 nm and 580 nm for the Ag- and Au-implanted samples, respectively. Corroborating the structural characterization results obtained by X-ray diffraction, these bands can be attributed to the localized LSPR frequencies of Ag and Au nanoparticles. According to the Mie theory for light scattering, these resonance frequencies depend on the size, shape and distance between nanoparticles [25,35–37], as well as on the dielectric medium in which they are embedded [25,38]. In particular, for Ag and Au nanoparticles, the localized surface plasmon resonance is typically expected to occur around 400 and 530 nm, respectively [14]. Borah et. al. [37] explored the behaviour of Au and Ag 20 nm nanoparticle arrays, concluding that decreasing the distance between NPs leads to an increase in the LSPR absorption's peak wavelength – for Ag from 400 to 600 nm and, for Au, from 550 to 700 nm. Therefore, the observed peaks are compatible with the SPR modes of Ag or Au nanoparticles, respectively.

Fig. 4a shows that the peak present at around 400 nm upon Ag implantation appears to undergo a redshift toward ~500 nm after annealing, which may indicate that the thermal treatment promotes nanoparticle growth or even a decrease of the average interparticle distance [25,37]. The growth of the nanoparticles is consistent with the fact that well defined-peaks assigned to Ag and Au were only observed via XRD after annealing. Although a more systematic investigation would be required to fully clarify this behaviour, the absence of a distinct absorption band in the as-implanted Au sample may suggest the lower diffusion and a larger nucleation energy of Au in $Ga_2O_3$ compared to Ag, thereby explaining the greater difficulty in forming Au nanoparticles during the implantation process.

## 3. Conclusions

In this work, we showed the successful formation of Ag and Au nanoparticles in (010)-oriented $\beta$-$Ga_2O_3$ single-crystals by ion implantation and annealing at 550 °C. Absorbance measurements revealed the presence of the surface plasmon resonance peak of such nanoparticles, corroborating the structural characterisation performed by X-ray diffraction. In particular, we showed that the formed nanoparticles are highly-ordered, displaying a well-defined crystallographic relation with respect to the matrix, namely $(0\bar{1}0)_\beta \parallel (110)_{Ag/Au}$ and $[102]_\beta \parallel [1\bar{1}2]_{Ag/Au}$. Notably, this is the same relation observed between monoclinic $\beta$-$Ga_2O_3$ and cubic $\gamma$-$Ga_2O_3$ induced by ion implantation. Moreover, the lattice constants of the $\gamma$-phase and of Ag/Au are commensurate, which may be crystallographically beneficial for the precipitation of the nanoparticles in the observed highly-ordered arrangement.

In short, this work reports, for the first time, the formation of Ag and Au nanoparticles in $\beta$-$Ga_2O_3$ single-crystals by ion implantation and thermal annealing. These nanoparticles presented a highly-ordered structure and a well-defined crystallographic relation with the substrate. Therefore, $Ga_2O_3$ can be considered a promising host with unique properties for the introduction of Au or Ag nanoparticles, opening the door for a wide number of future plasmonic-based applications such as multispectral UV-vis photodetectors.

## 4. Materials and methods

The $\beta$-$Ga_2O_3$ single-crystals used in this work were purchased from Novel Crystal Technology. These crystals were grown by the edge-defined film-fed growth method and cut yielding 5 mm × 5 mm samples with a thickness of ∼500 μm and (010) surface orientation.

The 110 keV $Ag^+$ and 160 keV $Au^+$ implantations were performed at room temperature at the 210 kV high flux ion implanter of the Laboratory of Accelerators of Instituto Superior Técnico (IST) [39] with a fluence of 5.0 × $10^{16}$ $cm^{-2}$, and a tilt angle of 10°. The SRIM Monte Carlo simulations [40] were performed in the full damage cascades calculation mode, considering displacement energies of 28 eV for Ga atoms and 14 eV for O atoms [41], as well as a density of 5.88 $g/cm^3$ for $\beta$-$Ga_2O_3$ [42].

The absorbance measurements were performed using a Shimadzu UV-1800 spectrophotometer at the Laboratory of Technological Assays in Clean Rooms of IST. This setup is based on two lamps — halogen for longer wavelengths and deuterium for shorter wavelengths and an integrated monochromator. The measured transmitted intensity of the light is automatically normalized to the light source spectrum, allowing for the absorbance to be directly obtained.

The X-ray Diffraction measurements were performed with the Bruker D8 Discover diffractometer of the Laboratory of Accelerators of IST. Two different configurations were employed in this work. In the high-resolution configuration, used in Fig. 2, the primary beam optics consists of a Göbel mirror, a 0.2 mm collimation slit, and a 2-bounce (220)-Ge monochromator, selecting the copper (Cu) K$\alpha_1$ X-ray line (wavelength of 1.5406 Å). The secondary beam path consisted of a 0.1 mm slit and a scintillation detector. While this geometry offers high angular resolution, the incident beam intensity is reduced, thus leading to poorer statistics. For this reason, the low-resolution configuration was employed for the remaining figures, where the 0.2 and 0.1 mm slits are substituted, respectively, by a 0.6 mm slit and long Soller slits, and the monochromator is removed.

**Acknowledgements**

The authors acknowledge funding of the Research Unit INESC MN from the Fundação para a Ciência e a Tecnologia (FCT) through Plurianual financing (UIDB/05367/2025, UID/PRR/5367/2025 and UID/PRR2/05367/2025) [doi: https://doi.org/10.54499/UIS/PRR/05367/2025 and https://doi.org/10.54499/UID/PRR2/05367/2025], as well as via the IonProGO project (2022.05329.PTDC, http://doi.org/10.54499/2022.05329.PTDC) and via the INESC MN Research Unit funding (UID/05367/2020) through Pluriannual BASE and PROGRAMATICO financing. D. M. Esteves (2022.09585.BD, https://doi.org/10.54499/2022.09585.BD) and A. S Sousa (2025.04778.BD) thank FCT for their PhD grants.